\documentclass[prd,nofootinbib,twocolumn]{revtex4}
\usepackage{graphics,color}
\usepackage{bm}
\usepackage{graphicx}
\usepackage{amsmath}
\usepackage{amssymb}
\usepackage{enumitem}
\usepackage{tensor}	
\usepackage[fleqn]{mathtools}
\usepackage{epstopdf}
\usepackage{hyperref}
\allowdisplaybreaks	
\usepackage[latin1]{inputenc}

\def\be{\begin{equation}}
\def\ee{\end{equation}}
\def\ba{\begin{eqnarray}}
\def\ea{\end{eqnarray}}
\def\l{\left}
\def\r{\right}
\def\f{\frac}

\def\nn{\nonumber \\}

\def\nn{\nonumber}

\newcommand{\lsim}{\raisebox{-0.13cm}{~\shortstack{$<$ \\[-0.07cm]
      $\sim$}}~}
\newcommand\lcdm{$\Lambda$CDM}
\def\eftcamb{\texttt{EFTCAMB}}

\begin{document}

\title{Cosmological constraints on Ho\v rava gravity revised in light of GW170817 and GRB170817A and the degeneracy with massive neutrinos}

\author{Noemi Frusciante$^1$, Micol Benetti$^{2,3}$}

\affiliation{ 
\smallskip
$^{1}$Instituto de Astrof\'isica e Ci\^encias do Espa\c{c}o, Faculdade de Ci\^encias da Universidade de Lisboa,  \\
 Edificio C8, Campo Grande, P-1749016, Lisboa, Portugal 
\smallskip \\
$^{2}$ Dipartimento di Fisica ``E. Pancini'', Universit\`a di Napoli ``Federico II'', Via Cinthia, I-80126, Napoli, Italy
\smallskip \\
$^{3}$Istituto Nazionale di Fisica Nucleare (INFN), sez. di Napoli, Via Cinthia 9, I-80126 Napoli, Italy
}

\begin{abstract}
We revise the cosmological bounds on Ho\v rava gravity taking into accounts the stringent constraint on the speed of propagation of gravitational waves from GW170817 and GRB170817A. In light of this we also investigate the  degeneracy between massive neutrinos and Ho\v rava gravity.  We show that a luminal propagation of gravitational waves suppresses the large-scale Cosmic Microwave Background (CMB) radiation temperature anisotropies and the presence of massive neutrinos increases this effect. On the contrary large neutrinos mass  can compensate the modifications induced by Ho\v rava gravity in the lensing, matter and primordial B-mode power spectra.  Another degeneracy is found, at theoretical level, between the tensor-to-scalar ratio $r$ and massive neutrinos as well as with the model's parameters. 
We analyze these effects using  CMB, supernovae type Ia (SNIa), galaxy clustering and weak gravitational lensing measurements and we show how such  degeneracies are removed.  We find that the model's parameters are constrained to be very close to their General Relativity limits and  we get a two orders of magnitude improved upper bound, with respect to the Big Bang Nucleosynthesis constraint, on the deviation of the effective gravitational constant from the Newtonian one. 
The deviance information criterion suggests that in Ho\v rava gravity  $\Sigma m_\nu>0$ is favored when  CMB data only are considered, while the joint analysis of all datasets  prefers zero neutrinos mass. 
\end{abstract}

\date{\today}

\maketitle

\section{Introduction}

 Theoretical \cite{Weinberg:1988cp,Martin:2012bt,Joyce:2014kja,Carroll:2000fy,Weinberg:2000yb,Padilla:2015aaa} and observational \cite{Riess:2019cxk,Wong:2019kwg,Delubac:2014aqe,Dawson:2012va,Abazajian:2008wr,Freedman:2019jwv,Yuan:2019npk,deJong:2015wca,Aghanim:2018eyx} issues are challenging the cosmological standard model or $\Lambda$-cold-dark-matter ($\Lambda$CDM). Alternative proposals usually include
 an additional dynamical scalar degree of freedom (dof) thus entering in the realm of modified gravity (MG) theories~\cite{Horndeski:1974wa,Fujii:2003pa,Deffayet:2009mn,Clifton:2011jh,Tsujikawa:2010zza,Gleyzes:2014dya,Joyce:2014kja,Koyama:2015vza,Langlois:2015cwa,Ferreira:2019xrr,Frusciante:2019xia}.  The additional dof can be  among others the result  of  breaking  the Lorentz Invariance (LI). Ho\v rava gravity~\cite{Horava:2008ih,Horava:2009uw} is a Lorentz Violating (LV) theory that breaks the LI by adding geometrical operators with higher order spatial derivatives to the action without including higher order time derivatives. The theory is then invariant under the more restricted foliation-preserving diffeomorphisms: $t\rightarrow \tilde{t}(t)$ and $x^i\rightarrow \tilde{x}^i(t,x^i)$, and it is power-counting renormalizable~\cite{Visser:2009fg,Visser:2009ys}.  As such it is a candidate for an ultraviolet completion of General Relativity (GR). The general action is characterized by a potential $V(g_{ij},N)$ that depends on  the spatial metric $g_{ij}$ and lapse function, $N$, of the Arnowitt-Deser-Misner (ADM) metric and their spatial derivatives. The power counting renormalizability allows the potential to contain only those operators which are at least a sixth order in spatial derivatives in a four-dimensional space-time. 

Different versions of Ho\v rava gravity correspond to various forms of the potential (see ref.~\cite{Sotiriou:2010wn} for a review).  One can impose the lapse function to be only a function of time, $N=N(t)$, obtaining the so called projectable version~\cite{Horava:2009uw}. On the contrary, if the lapse is a  function of both space and time one has the non-projectable version.  Another option is that of detailed balance  which requires the potential to be derived from a superpotential~\cite{Horava:2009uw}. Both the projectable and detailed balance versions limit the proliferation  of operators allowed by the symmetry of the theory but their assumption is not based on any fundamental principle and in some cases they can lead to instabilities and strong coupling at low energies~\cite{Sotiriou:2009bx,Charmousis:2009tc,Blas:2009yd,Wang:2009yz,Afshordi:2009tt,Koyama:2009hc,Sotiriou:2010wn,Vernieri:2011aa,Vernieri:2012ms,Vernieri:2015uma}. In the following we will consider the low-energy cosmology of the non-projectable version of the theory~\cite{Blas:2009qj} which is free from these pathologies  and shows a rich phenomenology compared to $\Lambda$CDM~\cite{Carroll:2004ai,Zuntz:2008zz,Kobayashi:2010eh,ArmendarizPicon:2010rs,Blas:2012vn,Audren:2013dwa,Audren:2014hza,Frusciante:2015maa,Munshi:2016lyb}. For instance, Ho\v rava gravity induces a rescaling of the gravitational constant at background level~\cite{Blas:2012vn}. This impacts on the relic abundance of  elementary particles in the Universe \cite{Carroll:2004ai} and  enhances  the growth of matter perturbations compared to $\Lambda$CDM~\cite{Kobayashi:2010eh,Zuntz:2008zz}. LV also induces modification in the Cosmic Microwave Background (CMB) power spectra through the lensing,  the Integrated Sachs-Wolfe (ISW) effects and a modified propagation of primordial Gravitational Waves (GWs)~\cite{Zuntz:2008zz,Audren:2014hza,Frusciante:2015maa,Gong:2018vbo}. 

Ho\v rava gravity is largely constrained by several probes which span from local tests to astrophysical and cosmological ones. These include: Big Bang Nucleosynthesis (BBN) bounds~\cite{Chen:2000xxa,Carroll:2004ai}; vacuum Cherenkov bounds, which exclude subluminal propagation for both tensor and scalar polarizations to a very high accuracy~\cite{Elliott:2005va}; post-Newtonian tests on the preferred-frame effects~\cite{Will:2014kxa,Bell:1995jz,Blas:2010hb,Blas:2011zd,Bonetti:2015oda,Frusciante:2015maa}; binary pulsars that can constrain the  modification on the orbital dynamics due to the emission of dipolar radiation~\cite{Yagi:2013ava}, cosmological data~\cite{Zuntz:2008zz,Audren:2014hza,Frusciante:2015maa} such as CMB, Baryon Acoustic Oscillations (BAO),  galaxy power spectrum, supernovae Ia (SNIa) measurements; and by the time delay  between the gamma-ray burst GRB170817A and the gravitational wave event GW170817~\cite{TheLIGOScientific:2017qsa,Monitor:2017mdv}. The latter sets a tight bound on the deviation of the speed of propagation of tensor modes, $c_t^2$, from the speed of light, $c$, of order $10^{-15}$. It implies that one of the free parameters of Ho\v rava gravity is found to be $\mathcal{O}(10^{-15} )$, leading to a revision of the  allowed parameter space~\cite{Gumrukcuoglu:2017ijh}. 

In this work we aim to revisit previous cosmological analysis on Ho\v rava gravity by considering the GWs bound and providing updated bounds. Previous cosmological analysis take into account  constraints from other sources (e.g. post-Newtonian tests, BBN, Cherenkov radiation) but not the tightest one from GWs. Thus as novelty we assume the GWs constraint in its stringent form, i.e. $c_t^2=1$ (in unit of $c=1$).  
Moreover we will extend previous works by including in the analysis massive neutrinos (with a varying mass) and   investigating the degeneracy between  Ho\v rava gravity and massive neutrinos. It is well known that MG models can mimic the effects of massive neutrinos on observables and impact the constraints on their mass~\cite{Motohashi:2012wc,He:2013qha,Baldi:2013iza,Hu:2014sea,Motohashi:2010sj,Bellomo:2016xhl,Alonso:2016suf,Frusciante:2019puu,Wright:2019qhf}.

The paper is organized as follows. In Sec.~\ref{sec:horava} we introduce  the low-energy action of Ho\v rava gravity and provide an overview of the current observational constraints on the model's parameters and stability relations. In Sec.~\ref{Sec:methodology} we outline the methodology adopted and introduce the formalism and the numerical tools used.  
In Sec.~\ref{Sec:degeneracy} we discuss the degeneracy between massive neutrinos and Ho\v rava gravity  by looking at the scalar angular  power spectra and matter power spectrum as well as the primordial B-mode spectrum. In Sec.~\ref{Sec:constraints} we present the cosmological constraints using the most updated datasets. Finally, we conclude in Sec.~\ref{Sec:conclusion}.

\section{Ho\v rava gravity}\label{sec:horava}

Let us consider the low-energy action of Ho\v rava gravity~\cite{Blas:2009qj} in the presence of matter fields, which can be written as follows
\begin{eqnarray} \label{horavaaction}
\mathcal{S}&=&\f{1}{16\pi G_H}\int{}d^4x\sqrt{-g}\left(K_{ij}K^{ij}-\lambda K^2 -2 \xi\bar{\Lambda} +\xi \mathcal{R}\r.\nn\\
&+&\l.\eta \,a_i a^i\right)+S_m[g_{\mu\nu},\chi_i],
\end{eqnarray}
where  $g_{\mu\nu}$ is the metric tensor and $g$ its determinant, $\mathcal{R}$ is the Ricci scalar of the three-dimensional space-like hypersurfaces, $K_{ij}$ is the extrinsic curvature, $K$ is its trace and  $a_i=\partial_i \mbox{ln} N$ is the 3-vector defined in terms of the the lapse function, $N$, of the ADM metric.  The three free parameters $\left\{\lambda,\xi,\eta\right\}$ are dimensionless running coupling constants and  $\bar{\Lambda}$ is the so called ``bare'' cosmological constant.  We define $S_m$ as the matter action for all matter fields, $\chi_i$. We further define $G_H=\xi \l(1-\f{\eta}{2\xi}\r)G_N$~\cite{Blas:2009qj} as  the coupling constant, where $G_N$ is the Newton gravitational constant. The  GR limit  is recovered when $\lambda=1$, $\xi=1$ and $\eta=0$.

Action~\eqref{horavaaction} propagates one scalar and two tensor  modes which have to satisfy some stability conditions. These require the avoidance of ghost instabilities and  positive speeds of propagation for both scalar and tensor modes which translate into the following requirements~\cite{Blas:2009qj}
\be
 0<\eta<2\xi\,, \qquad \lambda>1\,.
 \ee

Additional constraints on the model parameters can be found considering the bounds on the two parametrized post-Newtonian (PPN) parameters associated with the preferred frame effects, which  are $|\alpha_1|\lsim 3\cdot10^{-4}$ and $|\alpha_2|\lsim7\cdot 10^{-7}$ at 99.7\%C.L.~\cite{Will:2014kxa,Bell:1995jz}. These can be written in terms of the free parameters of the theory as follows~\cite{Blas:2010hb,Blas:2011zd,Bonetti:2015oda}:
\ba\label{PPN1}
&&\alpha_1=4(2\xi-\eta-2)\,,  \\
&&\alpha_2=-\frac{(\eta -2 \xi +2) (\eta  (2 \lambda -1)+\lambda  (3-4 \xi )+2 \xi -1)}{(\lambda -1) (\eta -2 \xi )} \,.\nn\\
&&\label{PPN2}
\ea
From that one can infer $\log_{10}(\lambda-1)<-4.1$ at 99.7\% C.L.~\cite{Frusciante:2015maa}. Usually, the bounds in Eqs.~\eqref{PPN1}-\eqref{PPN2} translate in 
\be\label{PPNrelation}
\eta=2(\xi-1)\,,
\ee
 and the parameter space reduces to a two dimensional plane.  
 
Assuming a flat Friedmann-Lema\^{i}tre-Robertson-Walker (FLRW) background  with line element
\be
ds^2=-dt^2+a(t)^2\delta_{ij}dx^idx^j \,,
\ee
where $a(t)$ is the scale factor and $\{t,x^i\}$ are respectively the time and spatial coordinates, the variation of the action~\eqref{horavaaction} with respect to the metric  provides the modified Friedmann equation which reads:
\be
H^2=\f{G_{c}}{G_N}H_0^2\l(\f{\Omega_m^0}{a^3}+\f{\Omega_r^0}{a^4}+\f{8\pi G_N}{3H_0^2}\rho_\nu+\Omega_{\rm DE}^0-1+\f{G_N}{G_c}\r),
\ee
where  $H\equiv\f{1}{a}\f{da}{dt}$ is the Hubble parameter and $H_0$ is its present time value; $\Omega_i^0\equiv8\pi G_N\rho_i^0/3H_0^2$ are the dimensionless density parameters and the subscript ``0'' stands for  their present day values, where $\rho_i$ stands for the density of baryons$+$cold dark matter (``m''), radiation (``r'') and massive neutrinos (``$\nu$'');  the dark energy (``DE'')  density parameter at present time, i.e. $\Omega_{\rm DE}^0$ , is defined from the flatness condition as follows~\cite{Frusciante:2015maa}
\be\label{DEtoday}
\Omega_{\rm DE}^0=\f{2\xi}{2\xi-\eta}\f{\bar{\Lambda}}{3H_0^2}+1-\f{3\lambda-1}{2\xi-\eta}\,.
\ee 
 This definition allows us to express $\bar{\Lambda}$ in terms of $\Omega_{DE}^0$ and to rewrite the  Friedmann equation only in terms of the parameters that will be sampled.
Additionally,  the effective gravitational constant is~\cite{Jacobson:2008aj,Blas:2009qj}
\be
G_{c}=\frac{ (\eta -2 \xi )}{1-3 \lambda }G_N\,.
\ee
The BBN constraint on the helium abundance~\cite{Izotov:2014fga,Aver:2015iza,Patrignani:2016xqp}  sets a bound on $G_{c}$ which is~\cite{Chen:2000xxa,Carroll:2004ai}:
\be
\l |\f{G_{c}}{G_N}-1\r|<\f{1}{8}\,,
\ee
and it can be used to further place bounds on the parameters of the theory.  
A combination of cosmological data such as the CMB, local Hubble measurements, SNIa,  galaxy power spectrum and BAO measurements 
set an improved upper limit on the deviation of the cosmological gravitational constant from the local Newtonian one~\cite{Frusciante:2015maa}, which is 
$G_{c}/G_N-1 < 0.028$ (at 99.7\% C.L.) and even stronger when  the PPN bounds are enforced, with  $G_{c}/G_N-1 < 6.1 \times 10^{-5}$ (99.7\% C.L.). 

The strongest constraint on the theory comes from  the joint observations of the  GW signal  from  a binary neutron star merger (GW170817)~\cite{TheLIGOScientific:2017qsa} and its gamma ray emission (GRB170817A)~\cite{Monitor:2017mdv}, which set a bound on the speed of propagation of tensor modes of  $-3\times 10^{-15} \leq c_t-1\leq 7 \times 10^{-16}$~\cite{Monitor:2017mdv}. In the case of Ho\v rava gravity  it implies $|\xi-1|\lsim10^{-15}$\,. The latter is several orders of magnitude stronger than the PPN bounds and as such it has been shown that the two dimensional plane identified by the relation in Eq.~\eqref{PPNrelation} has to be substituted with the  more informative two dimensional plane $\{\eta, \lambda\}$ characterized by  $\xi=1$~\cite{Gumrukcuoglu:2017ijh}.  In the present analysis we will only impose a priori the GWs bound in the form $\xi=1$  and we will not consider the condition in Eq.~\eqref{PPNrelation} in the following. In doing so we aim to investigate the power in constraining of cosmological datasets when compared to other bounds, specially those from Solar System.

Finally let us note that the bare cosmological constant $\bar{\Lambda}$ can be substituted with the dark energy  density parameter at present time in Eq.~\eqref{DEtoday}~\cite{Frusciante:2015maa}.  Therefore $\bar{\Lambda}$ will not be considered as a free parameter in the following analysis. 

\section{Methodology}\label{Sec:methodology}

The investigation of  Ho\v rava gravity at linear cosmological scales will be performed within the Effective Field theory (EFT) approach for dark energy and modified gravity~\cite{Gubitosi:2012hu,Bloomfield:2012ff,Gleyzes:2013ooa,Piazza:2013pua,Tsujikawa:2014mba,Frusciante:2019xia}, using the Einstein-Boltzmann code  \eftcamb \cite{Hu:2013twa,Raveri:2014cka,Hu:2014oga}. 
The EFT formalism describes the evolution of MG theories with one additional scalar dof both at 
 background and linear cosmological scales through a number of functions of time known as EFT functions. 
In this work we will  follow the methodology developed in Ref.~\cite{Frusciante:2015maa}, where the Ho\v rava gravity model has been implemented in \eftcamb, and we will use the resulting patch which is  publicly available~\footnote{Web page: \url{http://www.eftcamb.org}}.  

 The EFT action for Ho\v rava gravity with $c_t^2=1$, up to second order in perturbations, reads:

 \begin{align}
\mathcal{S}_{EFT} &= \int d^4x \sqrt{-g}  \bigg\{ \frac{m_0^2}{2} \left(1+\Omega \right)R + \Lambda(t) - c(t)\delta g^{00}  \nn \\
  &- \frac{c(t)}{4} \left( \delta g^{00} \right)^2- \frac{\bar{M}_2^2 }{2} \left( \delta K\right)^2\nn\\
      &+ m_2^2 h^{\mu\nu}\partial_{\mu}(g^{00})\partial_{\nu}(g^{00}) \bigg\} + S_{m} [g_{\mu \nu}, \chi_i],\label{actioneft}
\end{align}
where $m_0^2$ is the Planck mass and $R$ is the 4D Ricci scalar, $\delta g^{00}, \delta K$ are  the perturbations respectively of the upper time-time component of the metric and the trace  of the extrinsic curvature,   $h^{\mu\nu}=\left(g^{\mu\nu}+n^{\mu} n^{\nu}\right)$ is the induced metric with $n_\mu$ being  the unit vector perpendicular to the time slicing. $\Omega, c, \Lambda, \bar{M}_2^2, m_2^2$ are the EFT functions.  We note that $\Lambda$ and $c$ can be expressed in terms of $\Omega$, $H$ and the densities and pressures of matter fluids by using the background field equations, see Refs.~\cite{Gubitosi:2012hu,Bloomfield:2012ff} for details, and the remaining three EFT functions are~\cite{Frusciante:2015maa,Frusciante:2016xoj}:
\begin{align}\label{mapping}
&1+\Omega=\f{2}{(2-\eta)},  \\
&\bar{M}_2^2 =-2\f{m_0^2}{(2-\eta)}(1-\lambda),  \\
&m^2_2=\f{m_0^2 \eta}{4(2-\eta)}\,.
\end{align}
We refer the reader to Ref.~\cite{Frusciante:2015maa} for  further details about the background and linear perturbation equations implemented in \eftcamb.

The first part of our analysis will be the  study of the impact of massive neutrinos on the  cosmological observables and any degeneracy that might arise between massive neutrinos and the modifications of gravity  induced by LV. In detail, we list in Tab.~\ref{table:models} the values of the parameters for Ho\v rava gravity for the cases $H1$, $H2$ and $H3$ without massive neutrinos and $H1+\nu$, $H2+\nu$ and $H3+\nu$ with the summed neutrino mass $\sum m_\nu=0.85$ eV.    These  values are  bigger than the observational constraints we will present in Sec. \ref{Sec:constraints} and PPN bounds and  they serve only to visualize and quantify the modifications.  As a reference in our analysis we always include the \lcdm \, model. 

Finally, we will perform a Markov Chain Monte Carlo (MCMC) analysis using the \texttt{EFTCosmoMC} code~\cite{Raveri:2014cka} and the  datasets employed are listed in Sec.~\ref{Sec:data}.

\begin{center}
\begin{table}[t!]
\begin{tabular}{|l|c|c|c|}
\hline
Model &\,\,\, $\lambda-1$\,\,\,&\,\,\, $\eta$ \,\,\, & $\Sigma m_\nu$ (eV)  \\
\hline
\hline
H1         & 0.004 & 0.01  & --    \\
H1$+\nu$ &0.004 & 0.01 & 0.85   \\
H2& 0.04& 0.01 & --  \\
H2$+\nu$ & 0.04 & 0.01 & 0.85 \\
H3 & 0.004 & 0.1 & --\\
H3$+\nu$ &0.004 & 0.1 & 0.85  \\
\hline
\end{tabular}
\caption{Table with the values of  $\lambda$ and $\eta$ parameters for Ho\v rava gravity that we consider in Sec.~\ref{Sec:degeneracy}. We note that in this work $\xi=1$. Correspondingly we include also the cases with massive neutrinos.  The cosmological parameters are: $\Omega_b^0\,h^2=0.0226$, $\Omega_c^0\,h^2=0.112$  with $h=H_0/100$ and $H_0=70 \,\mbox{km}/\mbox{s}/\mbox{Mpc}$.  These cases study have been chosen to quantify the modification with respect to $\Lambda$CDM and the degeneracy with massive neutrinos. }
\label{table:models}
\end{table}
\end{center}

\section{Degeneracy between massive neutrinos and Ho\v rava gravity: a phenomenological description}\label{Sec:degeneracy}

Massive neutrinos have extended and measurable effects on the distribution of the large-scale structures,  the CMB  and the expansion history~\cite{Lesgourgues:2006nd,Wong:2011ip,Lattanzi:2016rre}. Their impact depends strictly on the value of their mass. The latest measured value of the summed neutrino mass from CMB Planck 2018 release sets the upper bound at $\Sigma m_\nu < 0.12$ eV (95\% C.L. with Planck TT,TE,EE+lowE +lensing+BAO) in the context of a flat standard cosmological model~\cite{Aghanim:2018eyx}, while the latest  direct measurement from  KATRIN experiment sets a higher upper limit of 1.1 eV  at 90\% C.L. \cite{Aker:2019uuj}.

In detail, massive neutrinos can change the height of the first acoustic peak of the CMB temperature-temperature power spectrum due to the early Integrated Sachs Wolfe (ISW) effect, suppress the weak lensing effect and dump the growth of structure on small scales~\cite{Lewis:2002nc}.
Similar effects are also characteristic of  DE and  MG models and as such a degeneracy between massive neutrinos and those models exists that strictly depends on the DE/MG model considered~\cite{Motohashi:2012wc,He:2013qha,Baldi:2013iza,Hu:2014sea,Motohashi:2010sj,Bellomo:2016xhl,Alonso:2016suf,Frusciante:2019puu,Wright:2019qhf}.  

In the following we show the imprint massive neutrinos leave on the dynamics of linear scalar and tensor perturbations in the context of Ho\v rava gravity and we investigate the degeneracy between massive neutrinos and the modified cosmological model under consideration.
To this purpose we also include the case without massive neutrinos and, for comparison, the $\Lambda$CDM model. For a complete overview of the cosmological effects of Lorentz violations we refer the reader to \cite{Carroll:2004ai,Zuntz:2008zz,Cai:2009dx,Kobayashi:2010eh,ArmendarizPicon:2010rs,Audren:2013dwa,Frusciante:2015maa,Munshi:2016lyb} and to \cite{Blas:2012vn,Audren:2014hza} for details about the effects of dark matter coupling with the aether.

 \begin{figure*}[t!]
\centering
\includegraphics[width=1.\textwidth]{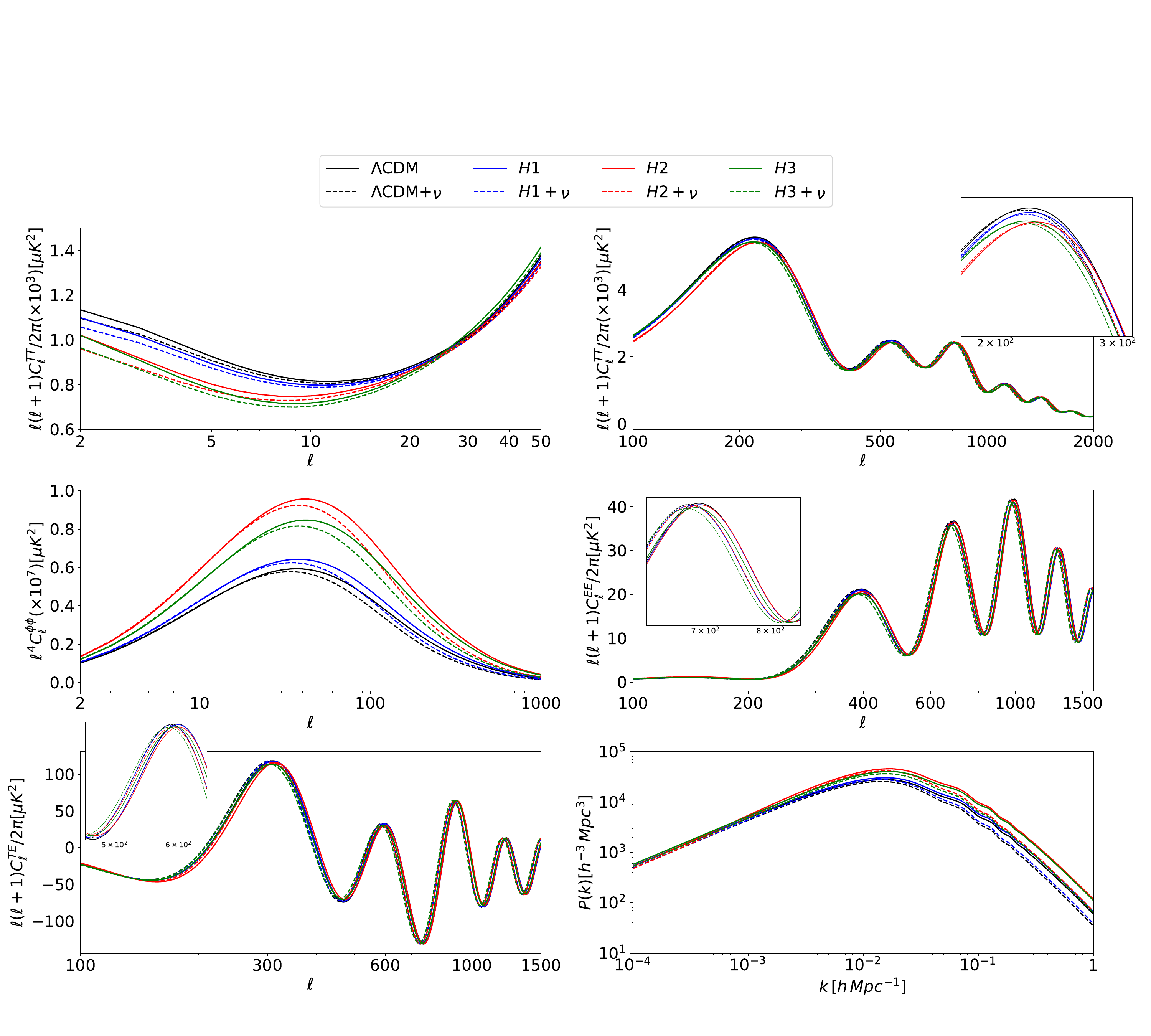}
\caption{\label{fig:spectranu} Power spectra of different cosmological observables for the Ho\v rava gravity models in Tab.~\ref{table:models} and $\Lambda$CDM. Top panels: CMB temperature-temperature power spectrum at low-$\ell$ (left) and high-$\ell$ (right). Central panels: lensing potential auto correlation power spectrum (left) and  E-modes power spectra (right).
Bottom panels:  cross power spectra of the temperature anisotropies and E-mode polarization (left) and matter power spectra (right).}
 \end{figure*}

\subsection{Scalar angular power spectra \& matter power spectrum}

We discuss the impact of  non-zero massive neutrino component on the scalar angular power spectra of CMB anisotropy and the matter power spectrum. The results are in Fig.~\ref{fig:spectranu}, where in top left panel we show the low-$\ell$ tail of the CMB temperature-temperature power spectrum. We note that Ho\v rava gravity models with a luminal propagation of GWs predict a suppressed ISW tail  for $\ell < 30$ with respect to $\Lambda$CDM which can be  up to 16\%. The $H1$ model is the closer one to $\Lambda$CDM, then there is the $H2$ characterized by a larger value of $\lambda$ (and same $\eta$) and finally the $H3$ which has the largest value of $\eta$ and the same value of $\lambda$ as in $H1$. This feature is due to the late time ISW effect, i.e. a modification of the time derivative of the lensing potential, $\dot{\Psi}+\dot{\Phi}$ (where $\Phi$ and $\Psi$ are the gravitational potentials).  In the specific case of Ho\v rava gravity, $\dot{\Psi}+\dot{\Phi}$ results to be enhanced  at late time with respect to $\Lambda$CDM.  We note that the MG effect goes in the same direction of those of  massive neutrinos. The latter indeed emphasizes the suppression. In $H1+\nu$, massive neutrinos  reduce the ISW tail of an additional $\sim 3.8\%$ with respect to the same model without massive neutrinos,  in $H2+\nu$  it is  $\sim 6\%$ and in $H3+\nu$ it is  $\sim 5.5\%$.  In $\Lambda$CDM$+\nu$ massive neutrinos also lower the low-$\ell$ tail with respect to the case without massive neutrinos of a factor up to $3.2\%$. Thus, in the case of Ho\v rava gravity the combined effects of massive neutrinos and modifications of gravity enhance the suppression. 
For $30<\ell<50$ the TT power spectra of $H1,H2$ ($+\nu$)    strictly follow $\Lambda$CDM or are slightly suppressed, while the one of $H3$  model is enhanced.  At these angular scales, the enhancement of $H3$ is lowered when massive neutrinos are included, compensating the MG effects. 

At high-$\ell$ in the TT power spectrum the MG effects are different than those of massive neutrinos. The former act on the height of the CMB peaks, e.g. we note a lower amplitude of the first and second peaks compared to $\Lambda$CDM for larger  values of $\lambda$ ($H2$) or larger value of $\eta$ ($H3$) due to a suppression of $\dot{\Psi}+\dot{\Phi}$ at early times. This suppression is more pronounced for the $H2$ case as modifications in the early ISW can be spotted already at $a \sim 10^{-3}$. The shift to higher multipoles in the position of the first two peaks is due to a different background expansion which is more pronounced in the $H2$ model having $G_c=0.94\,G_N$.  On the contrary, massive neutrinos  impact the position of the peaks by shifting the spectrum to lower multipoles for $\ell>200$, due to a change in the background expansion history.  Thus a non-zero neutrinos mass can compensate the shift to  higher $\ell$ in the CMB temperature anisotropy spectrum introduced by large value of the Ho\v rava gravity parameters.

Ho\v rava gravity models have an enhanced amplitude in the lensing power spectrum with respect to $\Lambda$CDM at all multipoles as shown in the left central panel in Fig.~\ref{fig:spectranu}. The deviation is larger for $H2$ ($\sim 90\%$), then it follows $H3$ ($\sim 75\%$) and finally $H1$ (12\%). Massive neutrinos as expected lower the amplitude  for $\ell>20$ and, as in the case of the TT power spectrum, the effect is larger for $H2+\nu$ and $H3+\nu$ compared to both $H1+\nu$ and $\Lambda$CDM$+\nu$. 

In the right central panel in Fig.~\ref{fig:spectranu} we show the EE-power spectrum. $H1$ model does not show any sizable effect due to MG with respect to $\Lambda$CDM.  A larger value of $\lambda$ ($H2$)  introduces an enhancement for $\ell <200$ which is  $\lesssim 20\%$ with respect to $\Lambda$CDM and then a suppression of the same order up to $\ell<500$. A larger value of $\eta$, as it is the case of $H3$, instead modifies the shape of the peaks and troughs for $\ell>400$ of about $10\%$. Massive neutrinos shift the overall spectrum to lower multipoles.  In the TE-power spectra, the effects of MG are present for $\ell<500$, see left bottom panel in Fig.~\ref{fig:spectranu}. These include both a shift of the position of the peaks to high-$\ell$ with respect to $\Lambda$CDM and in the height of peaks and troughs. The difference is larger for  $H2$ and $H3$ reflecting the effects in both the TT and EE power spectra. For the same reason massive neutrinos shift the spectrum to lower multipoles. The recent Planck data 2018 show an improved treatment on foregrounds and systematic effects on both TT and polarization spectra at high multipole, and also on EE spectra at low-$\ell$, which can help in constraining these effects~\cite{Akrami:2018odb,Aghanim:2019ame}.

Finally in the right bottom panel in Fig.~\ref{fig:spectranu}, we show the matter power spectrum. The latter is enhanced for all   Ho\v rava gravity models  with respect to $\Lambda$CDM. The larger deviation is for $H2$. While massive neutrinos suppress the growth of structures as expected~\cite{Lewis:2002nc}.  Thus the resulting effect is to mitigate the modifications due to large values of the Ho\v rava parameters.

  Let us stress that the large deviations with respect to $\Lambda$CDM  are a consequence of the  big values of the parameters we have selected for $H1$, $H2$ and $H3$. While these are larger than actual bounds these are very useful to  amplify the effect on the spectra and visualize the degeneracy.  Thus the  entity of the modifications that have been described in this section are specific to the choice of parameters but the overall directions of the modifications with respect to $\Lambda$CDM and impact of massive neutrinos hold for viable values of the parameters within the PPN and cosmological bounds.

 \begin{figure}[t!]
\centering
\includegraphics[width=0.49\textwidth]{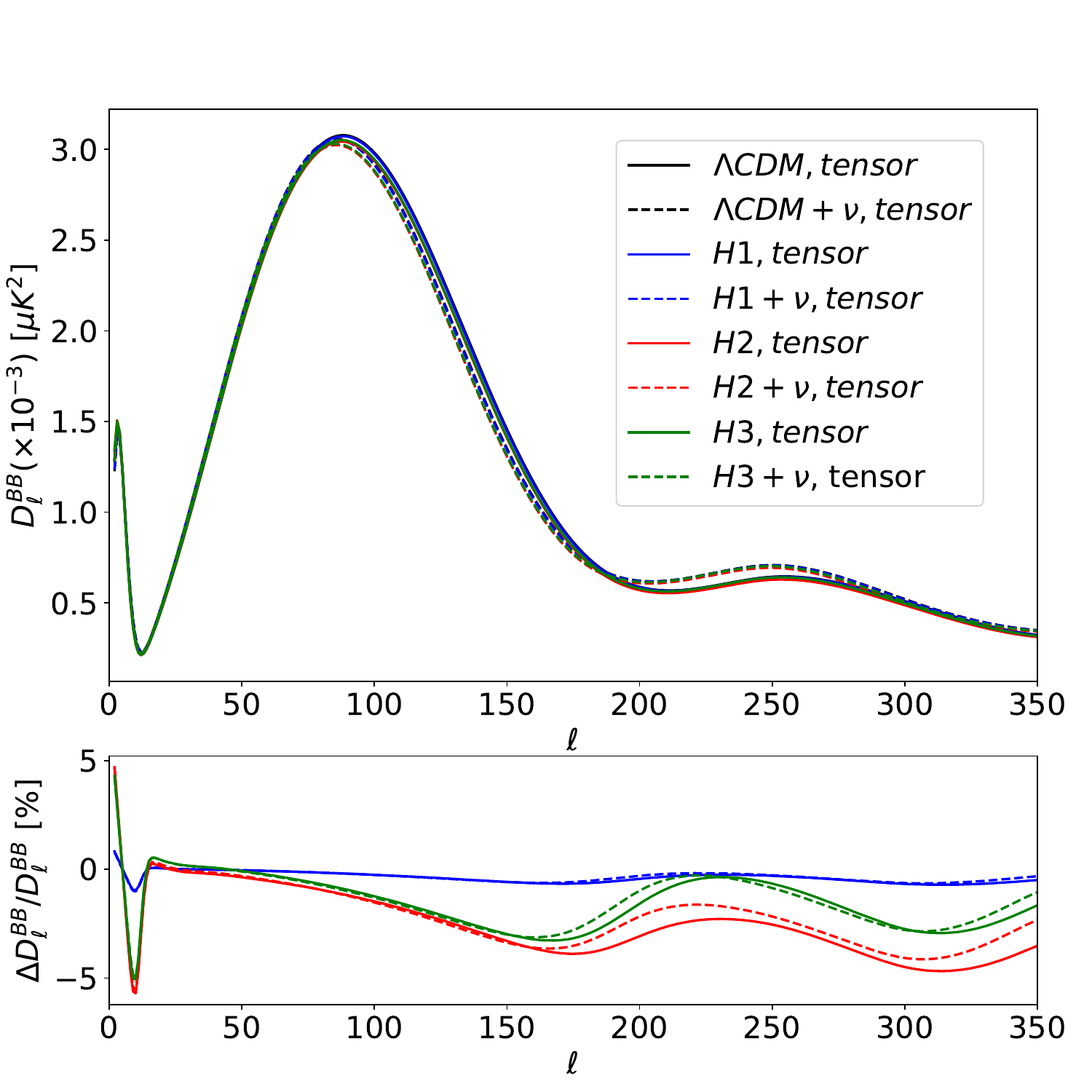}
\caption{\label{fig:spectraBBtens} The tensor contribution to the primordial BB power spectra for the Ho\v rava gravity models in Tab.~\ref{table:models} and $\Lambda$CDM.  We have set $r_{0.002}=0.05$. We define $D_\ell^{BB}=\ell(\ell+1)C_\ell^{BB}/2\pi$ and the relative difference as $\Delta D_\ell^{BB}/D_\ell^{BB}$, i.e the difference between the Ho\v rava gravity model and $\Lambda$CDM, divided by the standard cosmological model. }
 \end{figure}

 \begin{figure}[t!]
\centering
\includegraphics[width=0.49\textwidth]{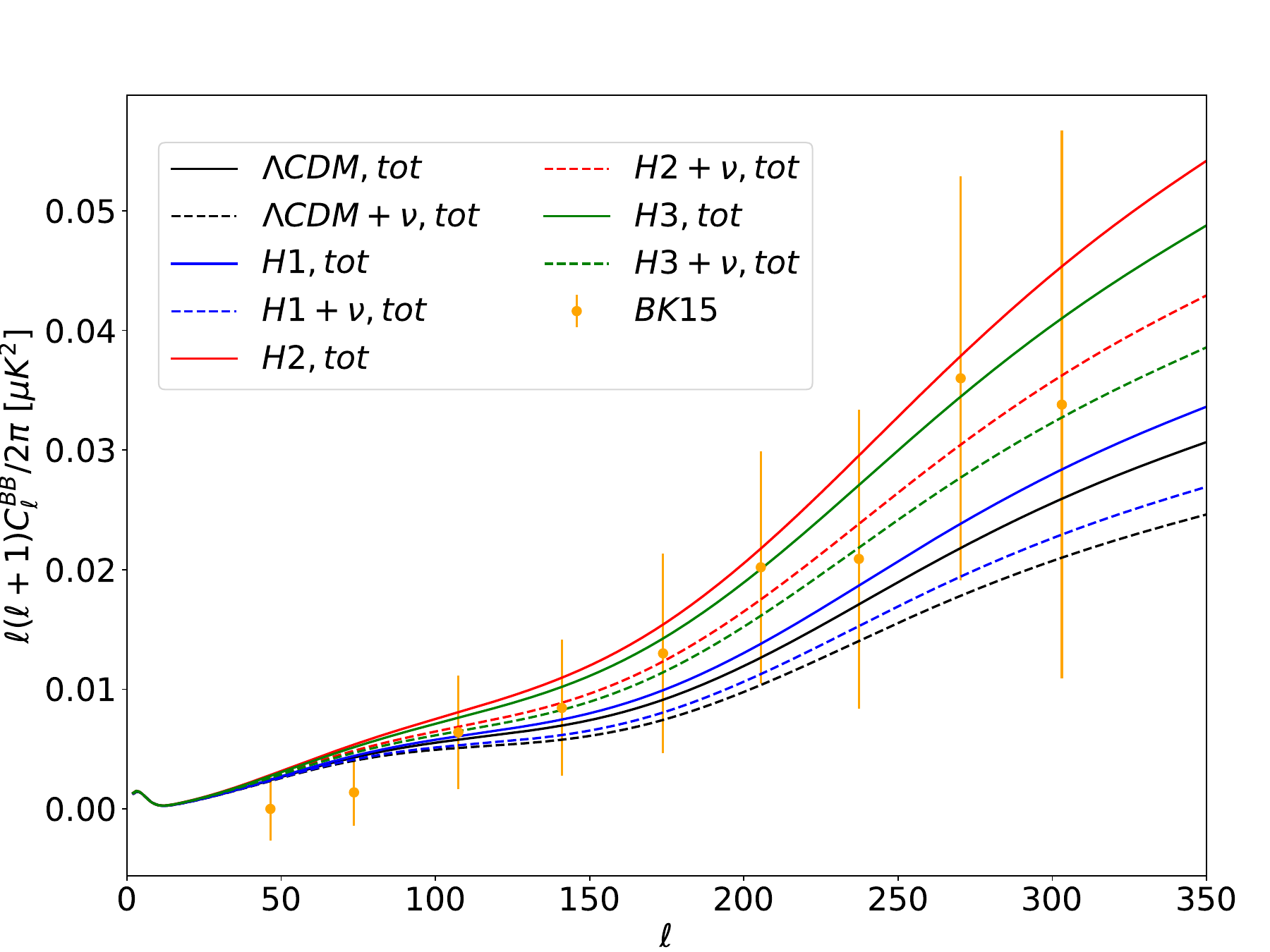}
\caption{\label{fig:spectraBBtot} The primordial  total BB spectra  including lensing for the Ho\v rava gravity models in Tab.~\ref{table:models} and $\Lambda$CDM. We also include the data points from BICEP2/Keck Array (BK15)~\cite{Ade:2018gkx}. We have set $r_{0.002}=0.05$.}
 \end{figure}

 \begin{figure}[t!]
\centering
\includegraphics[width=0.49\textwidth]{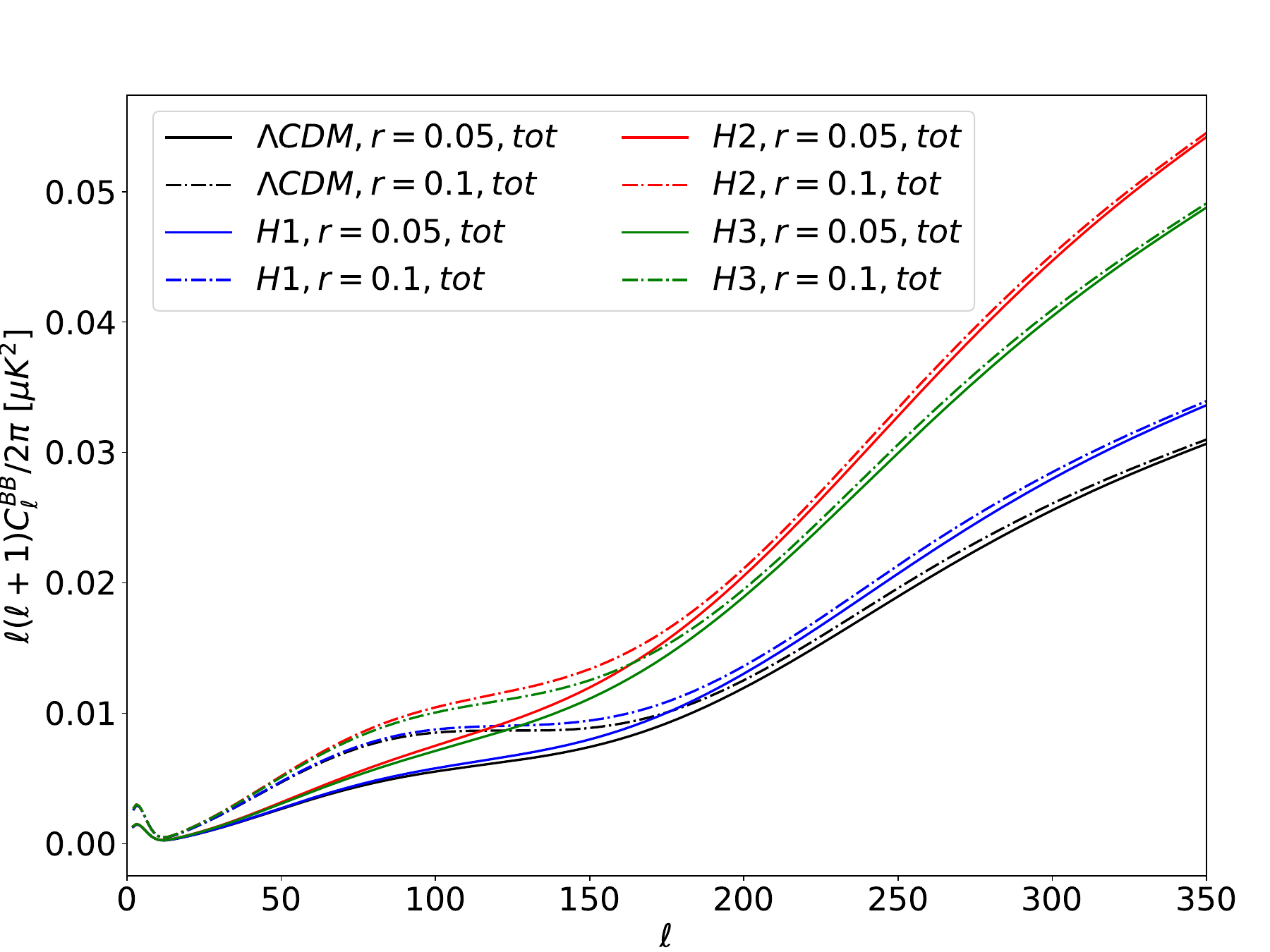}
\caption{\label{fig:spectraBBvaryr} The primordial  total BB spectra  including lensing for the Ho\v rava gravity models is Tab.~\ref{table:models} and $\Lambda$CDM. We show the impact of different values of the ratio of the tensor-to-scalar power spectra, $r$, on the total BB spectra.  We have chosen at the pivot scale, $k_*=0.002$ h/Mpc, two values for $r_{0.002}$: $r_{0.002}=0.05$ and $r_{0.002}=0.1$. }
 \end{figure}
\subsection{Primordial B-mode spectrum}\label{Sec:Bmode}

In this section we discuss the  Ho\v rava gravity phenomenology and that of massive neutrinos on the primordial B-spectrum of the CMB.

The  Ho\v rava gravity  evolution for tensor modes $h_{ij}^T$ with $\xi=1$ in Fourier space is given by the following equation:
\begin{align} \label{Eq:TensorEquation}
 \ddot{h}_{ij}^T +3H \dot{h}_{ij}^T + \f{k^2}{a^2} h_{ij}^T+  \frac{2-\eta}{2 m_0^2} \delta T _{ij}^T  = 0\,,
\end{align}
where dots are derivatives with respect to cosmic time and $\delta T_{ij}$ is the linear perturbation of the tensor component of anisotropic stress which contains the neutrinos and photons contribution.  The above equation is directly modified with respect to the one for $\Lambda$CDM because of the $2-\eta$ coefficient which regulates the  coupling between tensor modes and matter perturbations.  Let us also note that although the friction term, $3H$, is not directly modified, the evolution of the Hubble parameter in the Ho\v rava gravity model is rescaled by $G_c$ with respect to $\Lambda$CDM, thus affecting the amplitude of tensor modes.   The combination of these effects  leads to the features shown in Fig.~\ref{fig:spectraBBtens}. 
 While the BB-power spectrum for $H1$ mostly overlaps with the $\Lambda$CDM one, both the $H2$ and $H3$ models  show a general suppression of the peaks and troughs and a shift toward small-$\ell$. The overall differences are within $5\%$ and are larger for the $H2$ model because it has the smaller values of $G_c$ (for $H1$ $G_c=0.99 G_N$, for $H2$ $G_c=0.94 G_N$, for $H3$ $G_c=0.95 G_N$).  The inclusion of massive neutrinos further suppress the first peak and for larger multipoles ($\ell<300$) the BB-spectra are enhanced, a peculiar characteristic of massive neutrinos. They shift further the spectra toward smaller multipoles.  
  
The total spectra including lensing are shown in Fig.~\ref{fig:spectraBBtot}.  As already discussed in the previous section the lensing potential  is modified resulting in an enhancement of the BB-spectra for the Ho\v rava gravity models with respect to $\Lambda$CDM.  The inclusion of massive neutrinos  suppresses the tensor modes at  high-$\ell$, reducing the effects of MG.  We can infer that deviations due to large values of the Ho\v rava gravity parameters can be compensated by the inclusion of massive neutrinos. Thus,  in the BB-power spectrum  a degeneracy between massive neutrinos and the parameters of Ho\v rava gravity also exists. Furthermore, we notice that the modified total BB-spectra can accommodate  the BICEP2/Keck data points at high multipoles better than $\Lambda$CDM.  In particular, the case of $\Lambda$CDM seems to worsen the fit to data at small angular scales, even though it stays within the error. We will show in Sec.~\ref{Sec:results} that indeed this is the case. The joint analysis with CMB data shows a slightly better fit to data for Ho\v rava gravity with a non-zero neutrinos mass.  

Finally we  investigate the degeneracy between the tensor-to-scalar ratio $r$ and the Ho\v rava gravity parameters. $r$ has indeed been proven to be degenerate with modifications of gravity, as it is  the case of  modifications due to a non standard friction term~\cite{Pettorino:2014bka}.  In Fig.~\ref{fig:spectraBBvaryr}, we show the impact of this parameter on the total BB-power spectrum. Regardless of the cosmological model, changing the  value  of $r$ at the pivot scale $k_*=0.002$ h/Mpc from $r_{0.002}=0.05$ to $r_{0.002}=0.1$ leads to an overall enhancement of the total BB-power spectra at all angular scales. However the largest impact is for $\ell<150$.  Such enhancement is not only degenerate with  the parameters of Ho\v rava gravity as they can  also lead to a larger amplitude of the BB-power spectrum at these angular scales  but also with massive neutrinos. The latter indeed can compensate a larger value of $r$ as their effect is to dampt  the  BB-power spectrum amplitude. BICEP2/Keck data at low-$\ell$ can in principle  disentangle the degeneracy with $r$. 

\section{Cosmological Constraints}\label{Sec:constraints}

\subsection{Datasets}\label{Sec:data}

In the present analysis we consider  the following datasets:

\begin{itemize}

\item Measurements of the B-modes CMB power spectrum from the BICEP2 and Keck Array experiments  including the 2015 observing season~\cite{Ade:2018gkx} (hereafter ``BK15'');

\item Cosmic Microwave Background (CMB) measurements, through the Planck (2018) data~\cite{Aghanim:2019ame}, using ``TT,TE,EE+lowE" data by combination of temperature power spectra and cross correlation TE and EE over the range $\ell \in [30, 2508]$, the
low-$\ell$ temperature Commander likelihood, and the low-$\ell$ SimAll EE likelihood. We refer to this data set as ``Plk18";

\item The lensing reconstruction power spectrum from the latest Planck satellite data release (2018)~\cite{Aghanim:2019ame,Aghanim:2018oex}, hereafter indicated with ``lens";

\item Supernovae Type Ia data from the Joint Light-curve ``JLA" sample~\cite{Betoule:2014frx}, constructed from Supernova Legacy Survey (SNLS) and Sloan Digital SkySurvey (SDSS), and
consisting of 740 data points covering the redshift range $0.01< z <1.3$. It is worth mentioning that JLA sample, compared to other recent SNIa compilations, has the advantage of allowing the light-curve recalibration with the model under consideration, which is an important issue when testing alternative cosmologies~\cite{Taddei:2016iku,Benetti:2019lxu}.

\item Pantheon compilation~\cite{Scolnic:2017caz} of 1048 SNIa in the redshift range $0.01< z <2.3$. This is a larger sample than JLA that combines the subset of 276 newPan-STARRS1 SNIa with
useful distance estimates of SNIa from SNLS, SDSS, low-z and Hubble space telescope (HST) samples. It provides accurate relative luminosity distances. Hereafter we indicate this dataset with ``Pth";

\item  Dark Energy Survey Year-One (DES-1Y) results that combine galaxy clustering and weak gravitational lensing measurements, using 1321 square degrees of imaging data~\cite{Abbott:2017wau}.  We refer to this dataset as ``DES".

\end{itemize}

For the analysis we consider the following combinations:  BK15+Plk18, hereafter BKP, which will be the baseline dataset; on top of it we include first lens, DES and Pantheon (BKP+lens+DES+Pth), and then we consider JLA in place of Pantheon (BKP+lens+DES+JLA).

For the MCMC likelihood analysis  we use the \texttt{EFTCosmoMC} code~\cite{Raveri:2014cka}.
We consider the Ho\v rava gravity base model with a fixed $\xi=1$  to satisfy GWs constraints and varying $\lambda$ and $\eta$.  For the latter we consider flat priors: $\log_{10}(\lambda-1) \in {[-13, 0.1]}$ and $\log_{10} \eta \in {[-13, 0.1]}$.  The adopted ranges are consistent with stability conditions, which in any case are automatically enforced by the stability module of EFTCAMB \cite{Frusciante:2016xoj,DeFelice:2016ucp,Frusciante:2018vht}.  We use a logarithmic sampler for these parameters following Ref.~\cite{Frusciante:2015maa}. In addition to the model's parameters we vary the  physical densities of cold dark matter $\Omega_c h^2$ and baryons $\Omega_b h^2$, the angular size of the sound horizon at recombination $\theta_{MC}$, the reionization optical depth $\tau$, the primordial amplitude $\ln(10^{10} A_s)$ and spectral index $n_s$ of scalar perturbations and the tensor-to-scalar ratio $r$.  We also consider the additional case of a varying summed neutrino mass $\Sigma m_\nu$.


\begin{figure*}[t]
\centering
\includegraphics[width=1\textwidth]{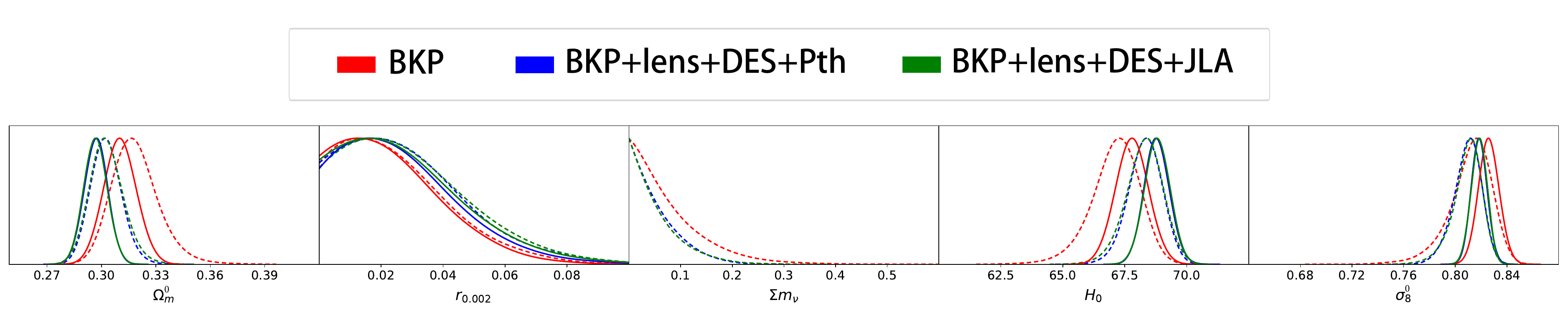}
\includegraphics[width=1\textwidth]{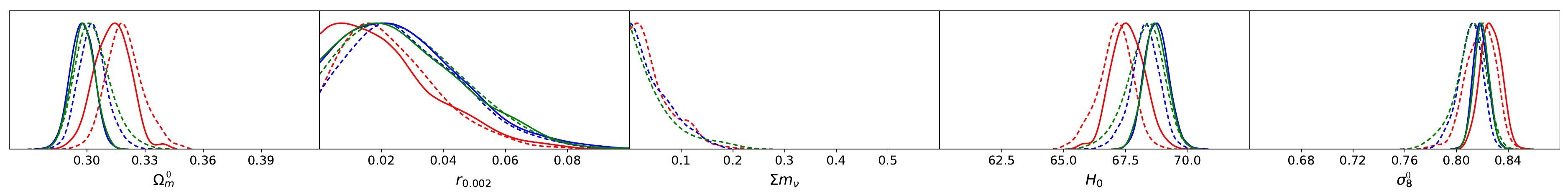}
\caption{\label{fig:1D_cosmo} Comparison between the $\Lambda$CDM (top panel) and Ho\v rava gravity (bottom panel) marginalized cosmological parameters. Solid lines indicate  the massless neutrino assumption, while dashed lines indicate the massive neutrinos extensions. The $68\%$ C.L.  are reported in Tab.~\ref{tab_bestfit_params}. }
 \end{figure*}


\renewcommand{\arraystretch}{1.3}
\begin{center}
\begin{table*}[t!]
    \begin{tabular}{| l | c | c | c | c | c |}
    \hline
   Model &  $\sigma_8^0$  &  $\Omega_m^0$ 
   & $H_0$  & $r_{0.002}$  & $\Sigma m_\nu$(eV)      
    	\\ \hline  \hline 
	$\Lambda$CDM (BKP)			& 
	$ 0.826 \pm 0.008$ 	        & 
	$ 0.310 \pm 0.008$ 	        & 
	$ 67.80 \pm 0.61$          &
	$ <0.054$                   &
	$ - $  \\
	$\Lambda$CDM (BKP+lens+DES+Pth)			&
	 $0.819 \pm 0.006  $ 	        &
	 $0.297 \pm 0.006  $           	& 
	 $68.80 \pm 0.47     $         &
	 $<0.061$	                   &
	$ - $    \\
	$\Lambda$CDM (BKP+lens+DES+JLA)	
	& $  0.819 \pm 0.006   $ 	        & 
	$ 0.297 \pm 0.006  $ 	           	& 
	$  68.81 \pm 0.47     $         &  
	$ <0.064$	         &
    $ - $   \\  \hline  
	$\Lambda$CDM+$\nu$ (BKP)			&
	 $  0.811 \pm 0.016 $ 	&
	 $  0.319 \pm 0.012$ 		&
	 $  67.11 \pm 0.91 $    &  
	 $ < 0.058 $	            &
	 $  <0.211 $  \\
      $\Lambda$CDM+$\nu$ (BKP+lens+DES+Pth)		& 
      $  0.809\pm 0.010 $ 	&
      $  0.303 \pm 0.008  $		&
      $  68.31 \pm 0.61   $    &
      $ <0.065$                           &
      $  <0.139  $  \\
	$\Lambda$CDM+$\nu$ (BKP+lens+DES+JLA)	&
   $ 0.809 \pm 0.010  $   	&
   $ 0.303 \pm0.008 $	&
   $ 68.28 \pm 0.68     $    & 
   $ <0.065$               & 
   $ <0.149 $  \\ 
   \hline  
   \hline
	Ho\v rava (BKP)		&
	 $  0.826 \pm 0.008  $ 	&
	 $  0.313 \pm 0.009 $	&
	 $ 67.59 \pm 0.64     $   &
	 $ <0.055 $	          &
	 $ - $  \\
        Ho\v rava(BKP+lens+DES+Pth) & 
       $  0.819 \pm 0.006  $ 	&
       $ 0.298 \pm 0.006 $		&
       $  68.74 \pm 0.47   $   &
       $ <0.063$	  &
       $ - $  \\
	 Ho\v rava (BKP+lens+DES+JLA)  & 
	 $  0.820 \pm 0.006  $ 	&  
	 $  0.298 \pm 0.006 $	& 
	 $  68.71 \pm 0.47   $   & 
	 $ <0.063 $	& 
	 $ - $  \\ \hline 
	 Ho\v rava+$\nu$ (BKP)   	& 
	 $  0.818 \pm 0.011 $ 	&  
	 $  0.319 \pm 0.009  $ 		& 
	 $  67.09 \pm 0.66   $    &  
	 $ <0.055 $	  &   
	 $  <0.125  $   \\
         Ho\v rava+$\nu$ (BKP+lens+DES+Pth)   	& 
         $  0.810 \pm 0.009 $ 	&  
         $  0.303 \pm 0.007 $	& 
         $  68.26 \pm 0.57    $    &  
         $ <0.060 $	&  
         $  <0.130 $   \\
         Ho\v rava+$\nu$  (BKP+lens+DES+JLA)  	& 
         $  0.810 \pm 0.011  $ 	         &  
         $  0.303 \pm 0.008 $		& 
         $  68.25 \pm 0.66    $         & 
         $ <0.060 $	  & 
         $ <0.165$   \\
	    \hline
     \end{tabular}
     \caption{\label{tab_bestfit_params} Marginalized constraints on cosmological  parameters at $68\%$ C.L., the upper limits are at $95\%$ C.L. }
 \end{table*}
\end{center}
\renewcommand{\arraystretch}{1}

  \begin{figure}[t!]
\centering
\includegraphics[width=0.46\textwidth]{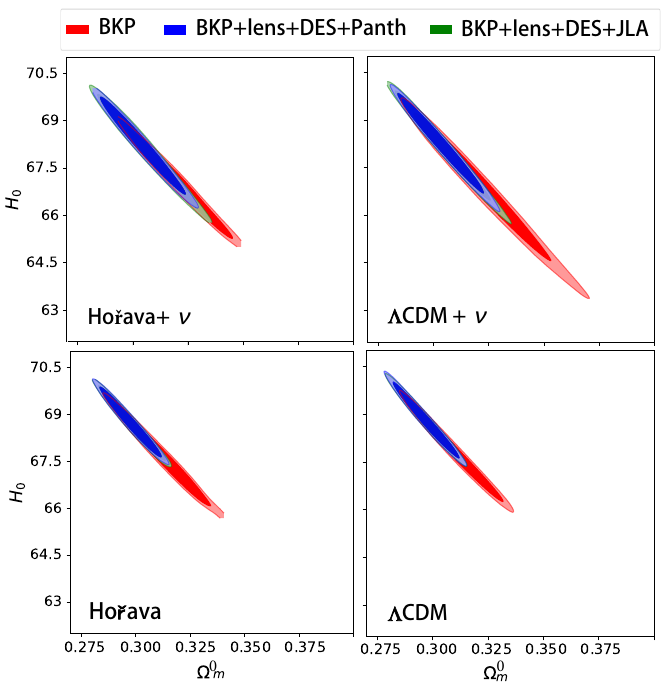}
\caption{\label{fig:2D_H0-Om}
 $H_0$-$\Omega_m^0$ plane for $\Lambda$CDM analysis (right panels) and Horava gravity model (left panels). }
 \end{figure}

  \begin{figure}[t!]
\centering
\includegraphics[width=0.46\textwidth]{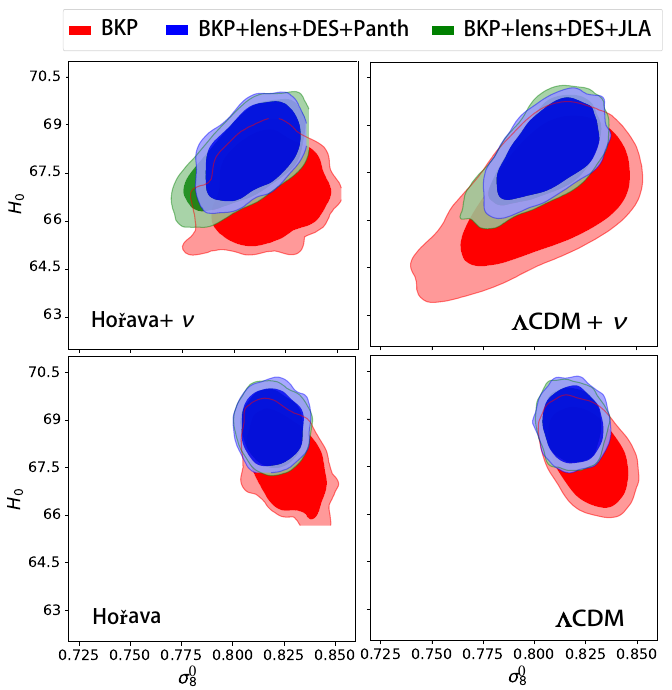}
\caption{\label{fig:2D_sigma8-H0} $H_0$-$\sigma_8^0$ plane for $\Lambda$CDM analysis (right panel) and Horava gravity model (left panel).}
 \end{figure}
 
 
\subsection{Results}\label{Sec:results}

This section is dedicated to the discussion of the cosmological and model parameters constraints of Ho\v rava gravity. We consider both the model with and without massive neutrinos. For reference we also include the results for $\Lambda$CDM in these two scenarios.  We present the results of a selection of  the cosmological parameters today $\{\Omega_m^0, H_0, \sigma_8^0, r_{0.002},\Sigma m_\nu\}$  in Tab.~\ref{tab_bestfit_params} at 68\% C.L. In Tab.~\ref{Tab:Constraints_Horava} we include the constraints on the model parameters, the derived constraints on $\alpha_1$, $\alpha_2$ and  the deviation of the effective gravitational constant, $G_{c}$, from $G_N$ at 68\% C.L.

 In Fig.~\ref{fig:1D_cosmo} we show the  marginalized likelihood of the cosmological parameters for $\Lambda$CDM (top panel) and Ho\v rava gravity (bottom panel). The cosmological parameters of Ho\v rava gravity  are consistent with those of the $\Lambda$CDM model (see Tab.~\ref{tab_bestfit_params}). In both models  the BKP dataset prefers a slightly larger central value for $\Omega_m^0$ with respect to the  other two combinations (BKP+lens+DES+Pth, BKP+lens+DES+JLA). Because of the anti-correlation  between $\Omega_m^0$ and $H_0$,   larger values of   $\Omega_m^0$  select a smaller values of $H_0$ and viceversa. We show this feature in Fig.~\ref{fig:2D_H0-Om}, where we see that the same  holds in the case massive neutrinos are included.  In the case of Ho\v rava$+\nu$ we note that the $\Omega_m^0$ upper limit (at 95\%C.L.) is slightly smaller with respect to $\Lambda$CDM$+\nu$ for the BKP dataset which in turn selects a higher lower limit for $H_0$.  
The anti-correlation also explains why $H_0$ goes toward smaller values when massive neutrinos are included. In this case indeed a larger value of $\Omega_m^0$ is expected. We also note that the extended datasets prefer lower central values of $\Omega_m^0$ (and higher values of $H_0$) in both cosmologies. 
We note that in the case of Ho\v rava$+\nu$ the dataset with JLA shows a higher upper bound for $\Omega_m^0$ ($<0.323$ at 95\% and smaller lower limit for $H_0>66.68$ at 95\%) with respect to the dataset with Pth ( $\Omega_m^0<0.318$ and $H_0>67.05$). The distinction between JLA and Pth is not present in $\Lambda$CDM. This is due to the fact that the Ho\v rava  posterior of massive neutrinos (see central bottom line in Fig.~\ref{fig:1D_cosmo} and Tab.~\ref{tab_bestfit_params}) for the dataset with JLA shows an higher upper limit with respect to $\Lambda$CDM. A similar consideration holds also for the baseline dataset, but  in this case the upper limit is smaller than the $\Lambda$CDM case as it is the upper bounds for massive neutrinos in the Ho\v rava gravity case.

Furthermore, in Fig.~\ref{fig:2D_sigma8-H0}  we show the marginalized 2D joint distribution for $H_0$ and $\sigma_8^0$. We note that the inclusion of massive neutrinos introduces a correlation between these two parameters, which is more pronounced in the standard cosmological model. We note that being the values of the cosmological parameters in Ho\v rava gravity  compatible with those of $\Lambda$CDM within the errors, Ho\v rava gravity suffers of the $H_0$~\cite{Riess:2019cxk,Delubac:2014aqe,Abbott:2017wau,Abbott:2017smn} and $\sigma_8^0$~\cite{Asgari:2019fkq} tensions which characterize the standard $\Lambda$CDM scenario. 

The bounds on the tensor-to-scalar ratio are the same in Ho\v rava gravity and $\Lambda$CDM independently on the presence of massive neutrinos.  The data analysis shows that the degeneracy between $r$, $\Sigma m_\nu$ and the Ho\v rava gravity parameters discussed in Sec.~\ref{Sec:Bmode} is removed. This is due to the fact that the modification introduced by varying  these parameters can go in the same direction or in the opposite one, depending on the observable considered. In some cases they affect a given cosmological observable in completely different ways, e.g. some shifting the power spectrum and others affecting its amplitude (see Sec.~\ref{Sec:degeneracy}). The datasets we chose are sensitive to different observables at different angular scales (lensing signal, T, E, B modes, galaxy clustering) in such a way that their combination is able to constrain these peculiar features and disentangle the degeneracies.

\begin{figure}[t!]
\centering
\includegraphics[width=0.4\textwidth]{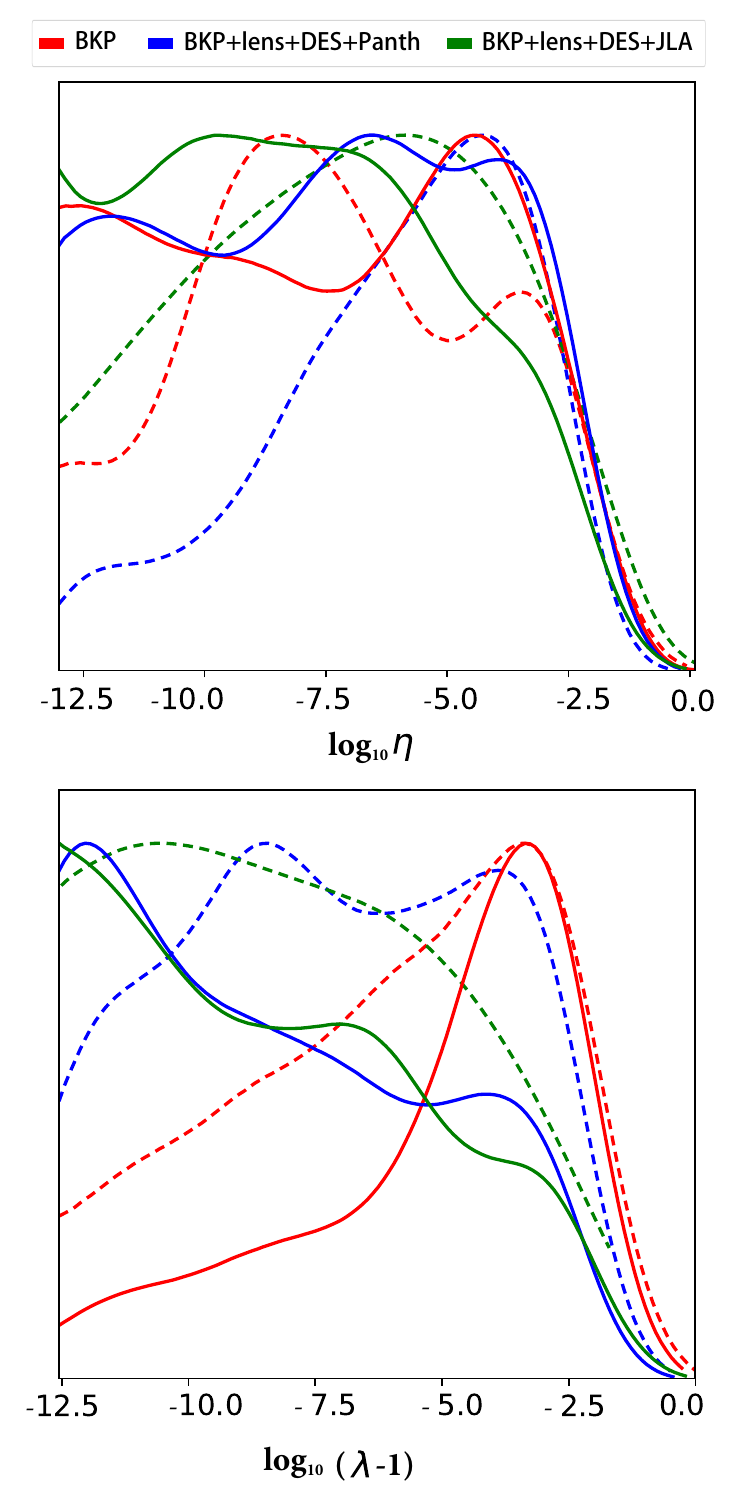}
\caption{\label{fig:1D_horava} The marginalized likelihood of $\log_{10} (\lambda-1)$ and $\log_{10}\eta$. Solid lines correspond to the case without massive neutrinos, dashed lines to the case with massive neutrinos.}
 \end{figure}

In Fig.~\ref{fig:1D_horava} we show the marginalized likelihood of the model parameters $\log_{10}(\lambda-1)$ and $\log_{10}\eta$ and the impact of the different combination of datasets.  We note that both parameters show a well defined upper limit at 95\% C.L.  
In the case of the baseline dataset, we note that $\lambda$ has peaked posteriors at 68\% C.L.: $\log_{10}(\lambda-1)=-2.6^{+0.1}_{-6.7}$ with massive neutrinos and $\log_{10}(\lambda-1)=-5.7 \pm 2.9$ without massive neutrinos. 
This is not the case for the posteriors of the other datasets.  
However, the upper limits in these cases are stringent: $\log_{10} (\lambda-1)<-3.2$ at 95\% C.L. for both  datasets without massive neutrinos and  $\log_{10} (\lambda-1)<-2.8$ at 95\% C.L with massive neutrinos. In top panel of Fig.~\ref{fig:1D_horava} we show the posterior of $\eta$. In the case  without massive neutrinos the datasets we considered are only able to set upper bounds, while when massive neutrinos are included it is also possible to obtain gaussian posteriors. In particular for the dataset with Pth we get  $\log_{10} \eta= -6.0^{+ 3.4}_{-1.6}$ at 68\% C.L.. 
In Tab.~\ref{Tab:Constraints_Horava}, we include the bounds on the PPN parameters and $G_c/G_N -1$. The derived constraints for $\alpha_1$ set a lower limit which is about one order weaker than the PPN bound. The latter, when $\xi=1$ can be read as a constraint on $\eta$: $\log_{10}\eta <-4.1$ at 99.7\% C.L. It is clear that such constraint is stronger than the ones we find using cosmological data (see Fig.~\ref{fig:1D_horava}). For $\alpha_2$ we find an upper bound which is several order of magnitude larger than the PPN constraint. Among the derived constraints on $\alpha_2$ the ones from BKP seem to be the stringent ones. That is because for this dataset the bounds on $\lambda$ include highest values. From Eq.~(\ref{PPN2}) we can deduce that a larger value of $\lambda$ decreases the estimation of $\alpha_2$, as already noted in Ref.~\cite{Frusciante:2015maa}. In this case  also we note that the PPN bound on $\log_{10} (\lambda-1)$ is stronger  than the cosmological one.  Additionally, we computed the bounds on the deviation of the effective gravitational constant from $G_N$ and we find that  in all cases considered they are two order of magnitude stronger than the BBN one.

We further  analyzed the case in which the additional PPN bunds are considered as prior and we find that the cosmological datasets used in this analysis do not show any improvement in the constraints. 

Finally, to determine whether  the Ho\v rava gravity model is favored with respect to $\Lambda$CDM, we use  the Deviance Information Criterion (DIC)~\cite{RSSB:RSSB12062}: 
\be
\text{DIC}:= \chi_\text{eff}^2 + 2 p_\text{D},
\ee
where $\chi_\text{eff}^2$ is the effective $\chi^2$ corresponding to the maximum likelihood and $p_\text{D} = \overline{\chi}_\text{eff}^2 - \chi_\text{eff}^2$. The bar stands for the average of the posterior distribution, and can be obtained from the output chains of the MCMC analysis. The maximum likelihood is computed employing the BOBYQA algorithm, implemented in \texttt{EFTCosmoMC} for likelihood maximization~\cite{Powell}. 
The DIC accounts for both the goodness of fit and the bayesian complexity of the model, or in other words takes into account its average performance (represented by the mean likelihood). The latter can also be considered a measure of the effective number of dofs in the model.

We then compute: 
\be
\Delta \text{DIC} = \text{DIC}_\text{Hor} - \text{DIC}_\text{$\Lambda$CDM}.
\ee
A negative value of $\Delta \text{DIC}$ means the Ho\v rava gravity model is supported by data over the $\Lambda$CDM one.
Let us stress that both the MCMC analysis and/or the minimization algorithm for the best fit introduce statistical noise and we must assume a scale to evaluate the $\Delta \text{DIC}$ high enough that any statistical fluke can be considered negligible when assessing the model selection criterion. Here we consider the convention based on the Jeffreys' scale for which  $\Delta \text{DIC}>10$ or $>5$ provide, respectively, strong/moderate evidence against the Ho\v rava gravity model. We compute also the $\Delta \text{DIC}$ between the Ho\v rava gravity model with and without massive neutrinos. The same Jeffreys' scale applies, where in this case positive values are against the presence of massive neutrinos.

We show the results in Tab.~\ref{Tab:Constraints_Horava}. We note that the $\Delta \text{DIC}$ values between Ho\v rava gravity and $\Lambda$CDM indicate generally a non-preference for a particular model, i.e. the data sets considered do not prefer one model over the other.
Even if without a significant statistical reading, is still of some interest the case of the analysis with the BKP data, for which the presence of massive neutrinos slows down the $\Delta\text{DIC}$ from 6.1 (moderate preference for the $\Lambda$CDM model) to $-0.4$.   Indeed this dataset seems to slightly favor the cosmological dynamics of Ho\v rava gravity  with massive neutrinos, however the evidence in support  of it is not sufficient to determine a proper preference between the models. 
 
 In conclusion the model selection analysis with the considered datasets does not give a definite conclusion for the preference of one model over the other.
 
 \renewcommand{\arraystretch}{0.8}
\begin{table*}[th!]
\centering
\begin{tabular}{|c|c|c|c|}
\hline
\multicolumn{4}{|c|}{Ho\v rava}\\
\hline
\hline
Parameters & 
BKP & 
BKP+lens+DES+Pth & 
BKP+lens+DES+JLA \\
\hline
\hline
$\log_{10} ({\lambda}-1)$  & 
$-5.7 \pm 2.9$ & 
$<-3.2$  & 
$<-3.1$  \\[1mm] 
$ \log_{10} \eta$ & 
$ <-2.8$ &  
$ <-2.7$  & 
$ <-2.9$  \\[1mm]  
\hline
\hline
$ \alpha_1 $ &  
$ > -0.008$ & 
$ > -0.008$ & 
$ > -0.010$ \\[1mm] 
$ \alpha_2 $  & 
$ <67 $ & 
$ <1.00 \times 10^6 $ & 
$ <2.85 \times 10^6$ \\[1mm] 
$ (G_{\rm c}/G_N -1)  $ &  
$ < 0.35 \times 10^{-2}$ & 
$ < 0.19 \times 10^{-2}$ & 
$ < 0.27 \times 10^{-2}$ \\[1mm] 
\hline
\hline
$ \Delta \text{DIC} $ &  
$ 6.1$ (moderate evidence)& 
$ 3.7$ (no evidence)&
$ 3.9$ (no evidence)\\
\hline
\hline
\multicolumn{4}{|c|}{Ho\v rava+$\nu$}\\
\hline
\hline
Parameters & BKP & BKP+lens+DES+Pth & BKP+lens+DES+JLA \\
\hline
$\log_{10} ({\lambda}-1)$  & 
$ -2.6 ^{+0.1}_{-6.7}$ &
$ <- 2.8$  & 
$ <-2.8$  \\[1mm] 
$ \log_{10} \eta$ & 
$ -6.9 ^{+3.6}_{-3.0}$ & 
$ -6.0^{+3.4}_{-1.6}$& 
$ -7.0^{+3.7}_{-3.0}$ \\[1mm] 
\hline
\hline
$ \alpha_1 $ &  
$ > -0.006$ & 
$ > -0.008$ & 
$ > -0.007$ \\[1mm] 
$ \alpha_2 $  & 
$ <0.18 \times 10^3 $& 
$ <0.48 \times 10^4$ & 
$ <0.28 \times 10^6$ \\[1mm] 
$ (G_{\rm c}/G_N -1)  $ &  
$ <0.44 \times 10^{-2}$ &
$ <0.22 \times 10^{-2}$ & 
$ <0.28 \times 10^{-2}$ \\[1mm] 
\hline
\hline
$ \Delta \text{DIC} $ &  
$ -0.4$ (no evidence)& 
$ 0.7$ (no evidence)&
$  4.1$ (no evidence)\\
\hline
\hline
$ \text{DIC}_{\text{Hor}+\nu} -\text{DIC}_\text{Hor} $ &  
$ -3.0$ (no evidence)& 
$ 0.5$ (no evidence) &
$ 3.1$ (no evidence)\\ 
\hline
\end{tabular}
\caption{The $68\%$ C.L. marginalized posterior bounds on the Ho\v rava and PPN parameters and the deviation of the effective gravitational constant from $G_N$. Upper limits indicated are at $95\%$ C.L.. We have also included the results for the $\Delta \text{DIC}$. }
\label{Tab:Constraints_Horava}
\end{table*}
\renewcommand{\arraystretch}{1}

\section{Conclusion}\label{Sec:conclusion}

We presented the phenomenology and observational constraints on the Ho\v rava gravity model in action~(\ref{horavaaction}) with $\xi=1$. This model is characterized by a luminal propagation of gravitational waves  in agreement with the GW170817 and GRB170817A  events. We performed a phenomenological analysis of scalar angular power spectra, matter power spectrum and primordial B-mode spectrum focusing on the degeneracy between modification of gravity and massive neutrinos. We find that both massive neutrinos and Ho\v rava gravity can suppress the ISW tail in the CMB TT power spectrum  with respect to $\Lambda$CDM. At the same time gravity modification enhances both the lensing  and matter power spectra while massive neutrinos mitigate these effects by suppressing the spectra amplitude. The same behavior  is  present in the total BB-spectrum also. In this case another degeneracy arises among Ho\v rava gravity parameters, massive neutrinos and the tensor-to-scalar ratio. Indeed large values of both $r$ and Ho\v rava gravity parameters can enhance the primordial total BB spectra, while a non-zero massive neutrinos component can suppress this feature.  The effects of a modified background evolution impact the high-$\ell$ TT power spectrum in different ways: by shifting the peaks to high multipoles and in the height of the CMB peaks which are suppressed due to an early ISW effect. Massive neutrinos instead shift the spectrum to lower multipoles. Thus a fine tuning among the mass of neutrinos and the values of Ho\v rava parameters can in principle compensate. The impact on the tensor BB power spectrum  are instead peculiar in the two cases: modification of gravity  suppresses  peaks and troughs while massive neutrinos further suppress the first peak but they enhance the spectrum for  larger multipoles. We used  CMB, SNIa,  galaxy clustering and weak gravitational lensing measurements in different combinations and we find that they were able to break these degeneracy due to the power in constraining different features of the model.

We provided observational constraints on model and cosmological parameters in the Ho\v rava gravity model using these data. We found that the cosmological parameters are compatible with those of $\Lambda$CDM in both scenarios (with/without massive neutrinos). As such the tensions in $H_0$ and $\sigma_8^0$ between low-redshift and CMB data are not alleviated in Ho\v rava gravity. The models parameters are severely constrained to be their GR limits. However their constraints are weaker than the ones obtained from the PPN bounds. We also computed the bounds on the deviation of the effective gravitational constant, $G_c$ from the Newtonian one $G_N$, and we found it to be two  order of magnitude stringent than the PPN one regardless of the dataset considered.

The model selection analysis using the Deviance Information Criterion (DIC)  suggests that CMB data from Planck 2018, BICEP2 and Keck Array experiments prefer in the case of Ho\v rava gravity $\Sigma m_\nu \neq 0$ ($\Delta  \text{DIC}=-3$), the opposite holds for the extended analysis, in particular for the combination of data including the JLA dataset. The CMB data are the solely which slightly prefer the Ho\v rava gravity model with massive neutrinos over the $\Lambda$CDM ($\Delta  \text{DIC}=-0.4$) even though without a significant statistical evidence; in all other cases (with/without massive neutrinos),  there is either a  mild preference for $\Lambda$CDM ($\Delta  \text{DIC}= 6.1$ for BKP without massive neutrinos, $\Delta  \text{DIC}= 4.1$ for BKP+lens+DES+JLA with massive neutrinos)  or a null preference. 

In conclusion the Ho\v rava gravity model can be still considered a viable candidate to explain the late time acceleration of the Universe and  it deserves further investigations particularly once new data will be available from next generation surveys, such as Euclid~\cite{Amendola:2016saw}, DESI~\cite{Levi:2019ggs},  LSST~\cite{Jha:2019rog}, SKA~\cite{Maartens:2015mra, Bacon:2018dui}, COrE~\cite{DiValentino:2016foa} and CMB-S4~\cite{Abazajian:2019tiv,Abazajian:2019eic}. These surveys will allow to measure cosmological/model parameters with unprecedented accuracy  and can help to definitely discriminate among the different cosmological models.

\acknowledgments

We thank Bin Hu and Daniele Vernieri for useful discussions and comments on the manuscript.
NF is  supported by Funda\c{c}\~{a}o para a  Ci\^{e}ncia e a Tecnologia (FCT) through the research grants UID/FIS/04434/2019, UIDB/04434/2020 and UIDP/04434/2020 and by FCT project ``DarkRipple -- Spacetime ripples in the dark gravitational Universe" with ref.~number PTDC/FIS-OUT/29048/2017.
MB acknowledge Istituto Nazionale di Fisica Nucleare (INFN), sezione di Napoli, iniziativa specifica QGSKY. This work was developed thanks to the High Performance Computing Center at the Universidade Federal do Rio Grande do Norte (NPAD/UFRN) and the National Observatory (ON) computational support. 
This paper is based upon work from COST Action (CANTATA/CA15117), supported by COST (European Cooperation in Science and Technology).

\bibliography{Horava}

\end{document}